\documentclass[aps,pra,twocolumn,showpacs,superscriptaddress]{revtex4-1}
\usepackage{graphicx}
\usepackage{dcolumn}
\usepackage{bm}
\usepackage{color}
\usepackage{times}
\usepackage{amssymb}
\usepackage{amsmath}
\usepackage{epsfig}
\usepackage{epstopdf}
\usepackage{dsfont}
\usepackage{subfigure}
\usepackage[colorlinks=true,citecolor=blue, linkcolor=blue,urlcolor=blue]{hyperref}
\usepackage{bookmark}
\usepackage{color}
\usepackage{rotating}
\usepackage[T1]{fontenc}

\definecolor{Green3}{rgb}{0.80,0.87,0.76}

\begin{document}
\renewcommand{\thefootnote}{\fnsymbol {footnote}}
	
	\title{\textbf{Entanglement and entropy uncertainty   in black hole quantum atmosphere}}
	
	\author{Shuai Zhang}
	\affiliation{School of Physics \& Optoelectronic Engineering, Anhui University, Hefei 230601, People's Republic of China}
\author{Li-Juan Li}
	\affiliation{School of Physics \& Optoelectronic Engineering, Anhui University, Hefei 230601, People's Republic of China}
\author{Xue-Ke Song}
	\affiliation{School of Physics \& Optoelectronic Engineering, Anhui University, Hefei 230601, People's Republic of China}

\author{Liu Ye}
	\affiliation{School of Physics \& Optoelectronic Engineering, Anhui University, Hefei 230601, People's Republic of China}

	\author{Dong Wang}
	\email{dwang@ahu.edu.cn}
    \affiliation{School of Physics \& Optoelectronic Engineering, Anhui University, Hefei 230601, People's Republic of China}
\begin{abstract}{
{In this work, we investigate the properties of Hawking radiation induced by the quantum atmosphere beyond the event horizon, by considering two detectors in Schwarzschild spacetime with the parameterized Hartle-Hawking temperature. \textcolor{black}{We explicitly study the dynamics of quantum entanglement and found that its characteristics are closely correlated with Hawking quantum radiation beyond the event horizon. Namely, its minimal value corresponds to the peak of Hawking radiation.} By virtue of the mutual information, we demonstrate the complementary relationship of the information distribution in the black hole. In addition, we detailedly discuss the influence of distance from the center of black hole to particle, radius of event horizon and Hartle-Hawking constant on the entropy uncertainty in the current scenario, and the results interestingly show that there exists an opposite correlation between the entanglement  and the entropy uncertainty. It is believed that our observation could provide a new perspective for understanding the black hole information paradox and black hole thermodynamics.}}
\end{abstract}
\date{\today}
\maketitle

\section{Introduction} 
The existence of black holes was first predicted by general relativity. In 2015, scientists successfully observed the transient gravitational wave signal for the first time through a detector \cite{1}, which is a breakthrough discovery that confirmed the existence of binary mass black hole systems. This is not only the first direct detection of gravitational waves in history,  also the first time we have observed a binary black hole merger. The event horizon of a black hole makes it impossible for any matter to return once it passes through. However, given quantum effects, the particles inside the black hole are
destined to gradually escape to the outside, resulting in Hawking radiation \cite{2}. This phenomenon is an intermediate bridge between quantum mechanics and gravity, and is at the heart of the information paradox of black hole \cite{3,4,5}.

Many achievements have been made in the study of black holes, however there are still some unsolved problems including the intrinsic characteristics of black holes. Through the examination of quantum fields within curved spacetime,  Hawking elucidated that the black holes are compelled to evaporate as a consequence of the semi-classical phenomenon of  particle creation \cite{6}. Hawking radiation is regarded as
the center behind black hole evaporation. Pairs of particles are created near a black hole's event horizon, causing the black hole to emit radiation and lose mass. The appearance of Hawking radiation has changed our understanding of what happens inside black holes. The origin of Hawking radiation in the evaporating black hole, where did Hawking radiation come from? In the previous study, a common view is that it is produced by excitation very close to the event horizon, and it is widely believed that the origin of Hawking radiation comes from quantum excitations in the near-horizon ($\Delta r=r-r_{{h}}\ll r_{{h}}$, where $r_{h}$ denotes the event horizon radius) \cite{5,6,7}. This perspective has bolstered the ``firewall'' argument and the claims regarding the significant role of UV-dependent  entanglement entropy in characterizing the quantum mechanics of black holes \cite{8}. However, through the total emission rate and stress tensor of Hawking radiation, Giddings finally concluded that the source of Hawking radiation is the near-horizon quantum region or quantum atmosphere, and the radial range of the near-horizon quantum region or ``quantum atmosphere'' is set by the horizon radius scale $(\Delta r=r_{A}-r_{h}\sim r_{h})$ \cite{8}.

Hawking radiation introduced quantum mechanics into the study of black holes, building a link between gravity and quantum mechanics. In the region of quantum mechanics,  uncertainty relation is deemed as one of backbones, which offers the precision limitation  with respect to two incompatible measurements. To be explicit, Heisenberg \cite{9} firstly proposed the uncertainty principle, stating that the position and momentum of a moving particle cannot be determined simultaneously. Since then, many scholars have conducted in-depth studies on uncertainty relations. Later, Kennard \cite{10} and Robertson \cite{11} presented an uncertainty relation based on standard deviations, which was generalized to apply to any pair of non-commuting observables and can be expressed as follows:
\begin{equation}
    \Delta \hat{X}\cdot\Delta \hat{R}\geq\frac{1}{2}\left|\langle[\hat{X},\hat{R}]\rangle\right|.
\end{equation}
Here, $\Delta \hat{X}$ represents the fluctuation of the observable $\hat{X}$, $\Delta \hat{X}=\sqrt{\langle \hat{X}^{2}\rangle-\langle \hat{X}\rangle^{2}}$, where $\langle \hat{X}\rangle=\mathrm{Tr}(\rho \hat{X})$ is the expected value of $\hat{X}$ in the state $\rho$, and $[\hat{X},\hat{R}]$ is the commutator. This relation depends on the state of the system. When the system's state is an eigenstate of the observable, the lower bound will disappear, resulting in a trivial result. In 1957, Everett \cite{Phys. 29. 454} and Hirschman \cite{Am. J. Math. 79. 152.} proposed the use of entropy to describe the uncertainty principle, which is known as the entropy uncertainty relation (EUR). Later, Bia{\l}ynicki-Birula and Mycielski \cite{Commun. Math. Phys. 44 129} provided a rigorous proof of the uncertainty relation between differential entropies of position and momentum.
Consequently, Deutsch proposed a new uncertainty relation via Shannon entropy \cite{12}. And then, Kraus \cite{13}, Maassen and Uffink \cite{14} refined this relation
into a concise form
\begin{equation}
    H(\hat{X})+H(\hat{R})\geq\log_2\frac{1}{c}=q_0,
\end{equation}
where $H(\hat{X})=-\sum_{i}p_{i}\mathrm{log}_2p_{i}$ represents the Shannon entropy of $\hat{X}$, $ p_{i}=\langle m_i| {\rho} |m_i\rangle $, $c=\max_{ij}\left|\left\langle m_{i}|n_{j}\right\rangle\right|^2$ denotes the maximum overlap between $\hat{X}$ and $\hat{R}$ with $|m_{i}\rangle$ and $|n_{j}\rangle$ being eigenstates of $\hat{X}$ and $\hat{R}$ respectively.

With respect to multipartite cases, the so-called quantum-memory-assisted entropy uncertainty relation (QMA-EUR) is proposed by Rene and Boileau \cite{16} and Berta {\it et al.}   \cite{15}, which is utilized
for measured uncertainty in bipartite or tripartite systems. And, the   QMA-EUR can be formulated as
\begin{equation}
    S(\hat{X}|B)+S\big(\hat{R}|B\big)\geq\log_2\frac{1}{c}+S(A|B),
\end{equation}
where $S(\hat{R}|B)=S(\rho_{\hat{R}B})-S(\rho_{B})$ refer to the conditional Von Neumann entropy of post-measurement state $\rho_{\hat{R}B}=\sum_{i}(|\mu_{i}\rangle_{A} \langle\mu_{i}|\otimes \mathbb{I})\rho_{AB}(|\mu_{i}\rangle_{A} \langle\mu_{i}|\otimes \mathbb{I})$ after particle $A$ is measured, with  $S(\rho)=-\sum_{i}\lambda_{i}\log_{2}\lambda_{i}$  and $\mathbb{I}$ being an identity operator in the Hilbert space \cite{17}.  Actually, the relation can be interpreted by the uncertainty game. Explicitly, suppose that there are two legitimate participators Alice and Bob in the game,
Bob in priori prepares a pair of   particles ($A$ and $B$),
and Bob sends particle $A$ to Alice, being entangled with his quantum memory $B$. Alice selects measurement $\hat{X}$ or $\hat{R}$ on system $A$  and attains the measuring result $\Theta$. Subsequently,
she informs Bob of his measurement choice via classical information, Bob can only win the game if he guesses $\Theta$ correctly, and the relation significantly provides the exactness of the guess.
Moreover, the EUR can be widely applied on many aspects in quantum information processing, including entanglement witness \cite{18,19,21,22}, quantum key distribution \cite{23,24,25}, EPR steering \cite{26,27}, quantum cryptography \cite{28,29}, quantum metrology \cite{30,31,32}. Recently, there are much progress on optimizing the EUR \cite{PhysRevLett.110.020402,PhysRevA.93.062123,PhysRevA.102.012206,Li2022,PhysRevA.104.062204, PhysRevE.106.054107,PhysRevA.106.062219,Phys. Lett. B 855 138876}.

On the other hand, \textcolor{black}{quantum entanglement is one of the important quantum resources,} which plays an irreplaceable and crucial role in the field of realistic quantum tasks. And it gives rise to many promising applications, such as quantum communication \cite{Theor. Comput. Sci.}, quantum computation \cite{Proc. R. Soc. Lond. A.} and quantum metrology \cite{40,41,42,43}. Therefore, it is crucial to understand genuine entanglement in the relativistic framework.  In particular, quantum entanglement influenced by Hawking radiation is a promising perspective to understand the information paradox of the  black hole \cite{44,6,5}. Motivated by this, we here focus on
uncovering the characteristics of entanglement redistribution and entropy uncertainty in the Schwarzschild black hole.


The Letter  is structured as follows. In Sec. \ref{2}, we briefly review the vacuum and excited states of the Dirac field under the background of Schwarzschild black hole. In Sec. \ref{3}, we offer the considered model and the Hartle-Hawking temperature in the presence of  quantum atmosphere. In Sec. \ref{4}, the quantum entanglement, entropy uncertainty and  relationship between both of them under quantum atmosphere are discussed in details. Finally, a concise summary is provided.

\section{The vacuum state and excited state in Schwarzschild spacetime}
\label{2}
The metric of the Schwarzschild black hole is denoted by
\textcolor{black}{\begin{align}
ds^{2} =
&\nonumber -\left(1-\frac{2M}{r}\right)dt^{2}+\left(1-\frac{2M}{r}\right)^{-1}dr^{2}\\
&          +r^{2}\left(d\theta^{2}+\sin^{2}\theta d\varphi^{2}\right),
\end{align}}
where $M$ is the mass of black hole \cite{47}. To facilitate our discussion, we shall adopt G (gravitational constant), c (speed of light), $\hbar$ (reduced Planck constant), $k_{B}$ (Boltzmann constant) as the fundamental units in our system. In Schwarzschild spacetime, the Dirac equation $[\gamma^{a}e_{a}^{\mu}(\partial_{\mu}+\Gamma_{\mu})]\psi=0$ for curved spacetime can be expressed as \cite{48}
\textcolor{black}{
\begin{align}
&\nonumber -\frac{\gamma_{0}}{\sqrt{1-\frac{2M}{r}}}\frac{\partial\psi}{\partial t}+\gamma_1\sqrt{1-\frac{2M}{r}} \left[\frac{\partial}{\partial r}+\frac{1}{r}+\frac{M}{2r(r-2M)}\right]\psi\\
&          +\frac{\gamma_{2}}{r}\left(\frac{\partial}{\partial\theta}+\frac{\cot\theta}{2}\right)\psi+\frac{\gamma_{3}}{r\sin\theta}\frac{\partial\psi}{\partial\varphi}=0,
\label{5}
\end{align}}
where $\gamma_{m}$ ($m$ = 0, 1, 2, 3) are the Dirac matrices, by solving the  Eq. (\ref{5}), the positive fermionic frequency output can be obtained as
\begin{equation}
   \psi_{k}^{\mathrm{I+}}=\xi e^{-i\omega u} ,
   \psi_{k}^{\mathrm{II+}}=\xi e^{i \omega u} ,
   \label{positive fermionic frequency output}
\end{equation}
where $\omega$ is the monochromatic frequency of the Dirac field, $k$ represents the field mode, $\xi$ is the 4-component Dirac spinor and $u=t-r_{*}$ and the tortoise coordinates are \textcolor{black}{$r_{*}=2(M-D) \ln \left[\frac{r-2 M}{2(M-D)}\right]+r$}. The positive frequency solutions of Eq. (\ref{positive fermionic frequency output}) correspond to the interior region and the exterior region of the event horizon, respectively. Using Damour and Ruffini method \cite{2}, the positive energy mode (Kruskal mode) is used to connect the two equations of Eq. (\ref{positive fermionic frequency output})
\begin{equation}
    \begin{aligned}&\Phi_{k, \text { I }}^{+}=e^{-2 \pi M \omega} \Psi_{-k, \text { II }}^{-}+e^{2 \pi M \omega} \Psi_{k, \text { I }}^{+},\\&
    \Phi_{k, \text { II }}^{+}=e^{-2 \pi M \omega} \Psi_{-k, \text { I }}^{-}+e^{2 \pi M \omega} \Psi_{k, \text { II }}^{+},\end{aligned}
\end{equation}
The Dirac field can also be expanded under the appropriate Kruskal modes, and can be obtained as
\begin{equation}
    \begin{aligned}&\psi=\int dk[2\cosh(4\pi M\omega_{i})]^{-\frac{1}{2}}[\hat{c}_{k}^{\text { II }}\Psi_{k,\text { II }}^{+}+\hat{d}_{-k}^{\text { II } \dagger}\Psi_{-k,\text { II }}^{-}\\&
    +\hat{c}_{k}^{\text { I }}\Psi_{k,\text { I }}^{+}+\hat{d}_{-k}^{\text { I }\dagger}\Psi_{-k,\text { I }}^{-}],\end{aligned}
\end{equation}
where the operators $\hat{c}_{k}$ and $\hat{d}_{-k}^{\dagger}$ serve as the creation and annihilation operators that operate on the Kruskal vacuum. According to the above discussion, it can be concluded that the Dirac field can be quantized by Schwarzschild and Kruskal modes. Using the Bogoliubov transform, the Kruskal vacuum state and excited state in Schwarzschild spacetime are expressed as
\begin{equation}
    \begin{aligned}&|0\rangle_{k}=\frac{1}{\sqrt{e^{-\frac{\omega_{i}}{T}}+1}}|0\rangle_{\text { I }}|0\rangle_{\text { II }}+\frac{1}{\sqrt{e^{\frac{\omega_{i}}{T}}+1}}|1\rangle_{\text { I }}|1\rangle_{\text { II }},\\&|1\rangle_{k}=|1\rangle_{\text { I }}|0\rangle_{\text { II }},\end{aligned}
    \label{the Kruskal vacuum state and excited state}
\end{equation}
where $T$ denotes Hawking temperature of the emitted radiation \cite{50}. For simplicity, we assume $\omega_{i} = \omega = 1$. $|x\rangle_{\text { I }}$ and $|x\rangle_{\text { II }}$ represent the fermionic modes outside the event horizon and the antifermionic modes inside the event horizon, respectively.

\section{Model}
\label{3}
For a $N$-qubit system with state of $X$-structure density matrix, it generally can be depicted as the form of
\begin{equation}
   \rho=\begin{pmatrix}
  A_{1}&  &  &  &  &  &  &C_{1} \\
  & A_{2}&  &  &  &  & C_{2} & \\
  &  &\ddots  &   &  &\begin{turn}{80}$\ddots$\end{turn}&  & \\
  &  &  & A_{n} & C_{n} &  &  & \\
  &  &  &  C_{n}^{*}& B_{n} &  &  & \\
  &  &\begin{turn}{80}$\ddots$\end{turn}  &  &  &\ddots  &  & \\
  &C_{2}^{*}  & &  &  &  & B_{2} & \\
  C_{1}^{*}&  &  &  &  &  &  &B_{1}
\end{pmatrix},
\end{equation}
where $n=2^{N-1}$. And $ {\textstyle \sum_{i,j=1}^n} (A_{i}+B_{j})=1$ is satisfied to guarantee the normalization of the matrix. In this case, the genuine concurrence for the $N$-partite $X$-state
can be expressed by
\begin{equation}
    C(\rho_{})=2\operatorname*{max}\{0,|C_{i}|-\nu_{i}\},i=1,\ldots,n,
\end{equation}
where $\nu_{i}= {\textstyle \sum_{j\ne i}^{n}} \sqrt{A_{j}B_{j}} $ \cite{51}.

To probe the quantumness of black hole quantum atmosphere, there are a pair of modes $A$ and $B$ in priori prepared in  state of Bell-type state
\begin{equation}
    \left|\psi \right\rangle_{AB}=\cos\alpha \left|00\right\rangle+\sin\alpha \left|11\right\rangle,
\end{equation}
where state's parameter \textcolor{black}{$\alpha\in [0,\pi/2 ]$}, and the mode $A$ and the mode $B$ are observed by the observers, say Alice and Bob. First, Alice and Bob remain in an asymptotically flat region. Then, Alice is left in the asymptotically flat region, and Bob falls towards the Schwarzschild black hole freely. As a result, the mode $B$ will experience the effects of Hawking radiation in the quantum atmosphere region, which is extended beyond the radius $r_{h}$ of an event horizon.

From Eq. (\ref{the Kruskal vacuum state and excited state}), it is observed that the transition from Kruskal modes to Schwarzschild modes corresponds to the transformations of two-mode squeezing. Under the above transformation, one can obtain
\begin{equation}
    \begin{aligned}&\left|\psi \right\rangle_{AB_{\text{I}}B_{\text{II}}}=\frac{\cos\alpha }{\sqrt{e^{\frac{-\omega_{i}  }{T} } +1} } \left|000\right\rangle\\&
    +\frac{\cos\alpha }{\sqrt{e^{\frac{\omega_{i}  }{T} } +1} } \left|011\right\rangle+ \sin\alpha \left|110\right\rangle.\end{aligned}
\end{equation}

Given the disconnection between regions I and II, and the fact that Bob cannot access the modes within the event horizon, mode $B_{\text{I}}$ located outside the event horizon are referred to as physically accessible, whereas mode $B_{\text{II}}$ situated inside the event horizon are designated as inaccessible. Through tracing over all degrees of freedom in region II, the state of the physically accessible subsystem I can be captured, thereby leading to   the reduced  density  matrix $\rho_{AB_{\text{I}}}$,   which is rewritten as follows:
\begin{equation}
    \rho_{AB_{\text{I}}}=\begin{pmatrix}
  \rho _{11} &0&0&\rho _{14}  \\
  0&\rho _{22} &0&0 \\
   0&0&0&0 \\
   \rho _{41} &0&0&\rho _{44} \\
\end{pmatrix},
\end{equation}
with $\rho _{11}=\frac{\cos^{2}\alpha}{e^{\frac{-\omega_{i}  }{T} } +1},\ \rho _{14}=\rho _{41}=\frac{\cos\alpha \sin\alpha}{\sqrt{e^{\frac{-\omega_{i}  }{T} } +1}},
    \ \rho _{22}=\frac{\cos^{2}\alpha}{e^{\frac{\omega_{i}  }{T} } +1},\ \rho _{44}=\sin^{2}\alpha$. As a result, the concurrence of $\rho_{AB_{\text{I}}}$ in physically accessible regions can be calculated as follows:
\begin{equation}
    C(\rho_{AB_{\text{I}}})=\frac{2 \cos\alpha \sin\alpha }{\sqrt{e^{\frac{-\omega_{i}  }{T} } +1} }.
    \label{concurrence1}
\end{equation}
Similarly, the concurrence of $\rho_{AB_{\text{II}}}$ can be given by
\begin{equation}
    C(\rho_{AB_{\text{II}}})=\frac{2 \cos\alpha \sin\alpha }{\sqrt{e^{\frac{\omega_{i}  }{T} } +1} }.
    \label{concurrence2}
\end{equation}

To investigate the potential characteristics of the black hole's quantum atmosphere, we can replace the hawking temperature $T$ in Eqs. (\ref{concurrence1}) and  (\ref{concurrence2}) with the local temperature $T_{HH}$ in Hartle-Hawking vacuum, which can be expressed as
\textcolor{black}{\begin{equation}
    \begin{aligned}&T_{HH}=T_{H}\sqrt{1-\frac{r_{h}}{r}}\\&
    \sqrt{1+2\frac{r_{h}}{r}+\left(\frac{r_{h}}{r}\right)^{2}\left(9+D_{HH}+36\ln\left(\frac{r_{h}}{r}\right)\right)},
    \label{local temperature}\end{aligned}
\end{equation}}
where \textcolor{black}{$T_H=1/(4\pi r_h)$} \cite{52,56,57}, and $D_{HH}$ represents the  constant in the stress tensor associated with the Hartle-Hawking vacuum. The constant $D_{HH}$ in Eq. (\ref{local temperature}) is considered to be arbitrary,
and it is not sufficient to use Hartle-Hawking's boundary conditions to fix $D_{HH}$, thereby some additional conditions are required to determine it. To avoid the occurrence of imaginary temperature in a region outside the horizon and to prevent temperatures that are inversely related to distance, the temperature remains real throughout the entire region and decreases monotonically as $r$ increases, for $D_{HH} \ge D_{c}\simeq 23.03$, then reaching a maximum value at $r_{c}\simeq 1.43r_{h}$. In what follows,  $D_{HH} \ge D_{c}$ is satisfied \cite{52}. Besides, from Eq. (\ref{local temperature}), it can be directly seen that the local temperature vanishes at the horizon ($r=r_{h}$), and  approaches the expected Hawking temperature as $r$ tends to infinity ($r\rightarrow \infty $).

Since the uncertainty of two incompatible Pauli measurements (${\hat{X}},{\hat{Z}}$) in ${\rho}_{AB_{\text{I}}}$ can be calculated as follows
\begin{equation}
U=S({\rho}_{\hat{X}B_{\text{I}}})+S({\rho}_{\hat{Z}B_{\text{I}}})-2S(\rho_{B_{\text{I}}}),
\end{equation}
where, the post-measurement states can be given by
\begin{equation}
    \begin{gathered}
{\rho}_{\hat{X}B_{\text{I}}} =\sum_{i}\left(|x_{i}\rangle\langle x_{i}|\otimes{\mathbb{I}}\right){\rho}_{AB_{\text{I}}}\left(|x_{i}\rangle\langle x_{i}|\otimes{\mathbb{I}}\right), \\
{\rho}_{\hat{Z}B_{\text{I}}} =\sum_{i}\left(|z_{i}\rangle\langle z_{i}|\otimes{\mathbb{I}}\right){\rho}_{AB_{\text{I}}}\left(|z_{i}\rangle\langle z_{i}|\otimes{\mathbb{I}}\right),
\end{gathered}
\end{equation}
and the reduced density matrix of its subsystem $B_{\text{I}}(A)$ can be represented as $\rho_{B_{\text{I}}(A)}=\operatorname{tr}_{A(B_{\text{I}})} \left(\rho_{A B_{\text{I}}}\right)$.

Besides, the mutual information is regarded as an important concept for quantifying the total correlation of a composite system, including quantum and classical parts, which can be mathematically expressed as
\begin{equation}
I\left(\rho_{A B_{\text{I}}}\right)=S\left(\rho_{A}\right)+S\left(\rho_{B_{\text{I}}}\right)-S\left(\rho_{A B_{\text{I}}}\right).
\end{equation}

\begin{figure}[t]
    \centering
    \includegraphics[width=8cm]{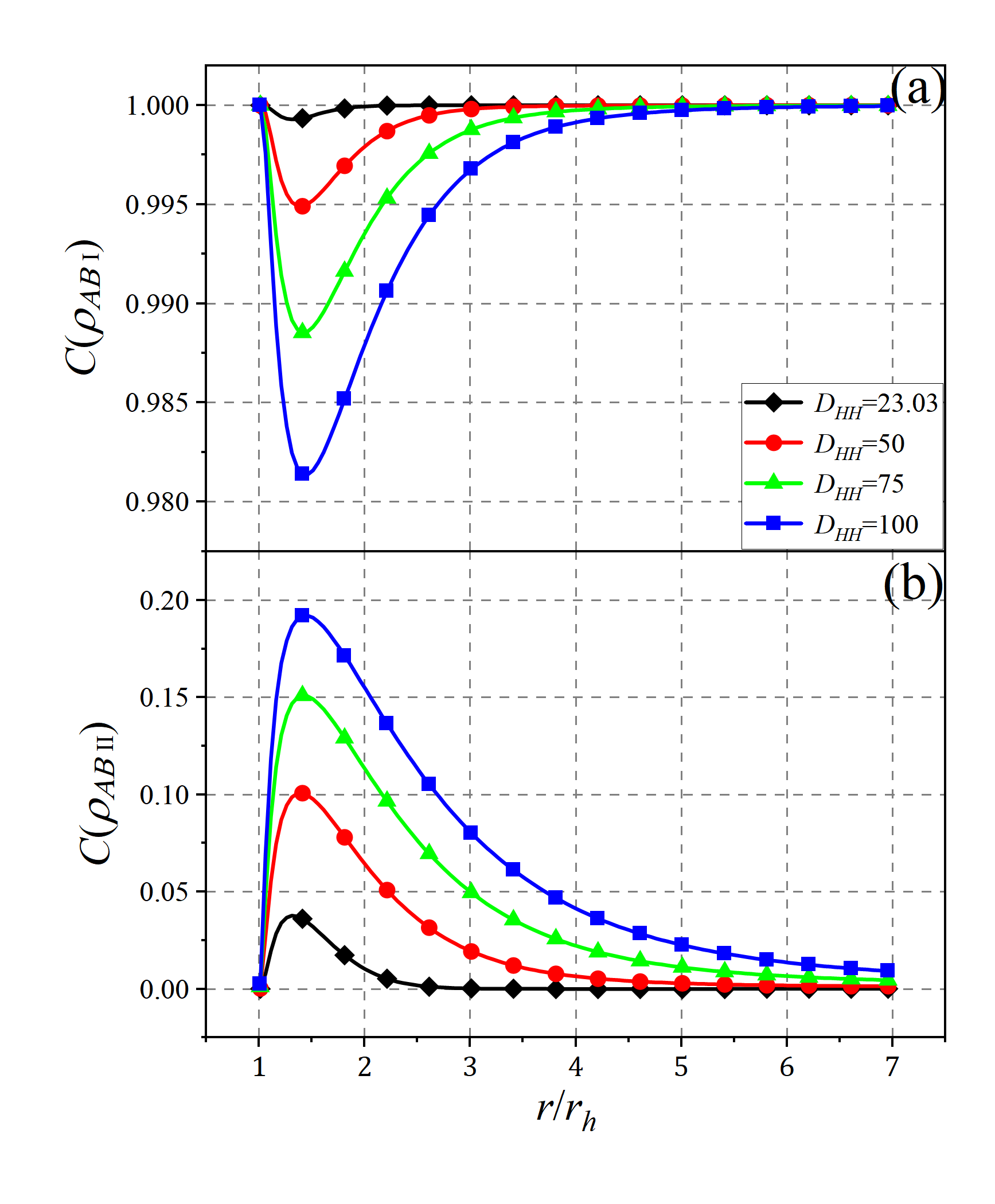}
    \caption{The system's entanglements as functions of the normalized distance $r/r_{h}$ for the different constant $D_{HH}$.
     Graph (a): the entanglement of $\rho_{AB_{\text{I}}}$ in the physically accessible region $C(\rho_{AB_{\text{I}}})$, and  Graph (b): the entanglement of $\rho_{AB_{\text{II}}}$ in the physically inaccessible region $C(\rho_{AB_{\text{I}}})$. The state's parameter $\alpha= {\pi}/{4}$ is set in the plots. }
    \label{fig1}
\end{figure}

\section{Thermal evolution of Concurrence and Entropy Uncertainty}
\label{4}

 Fig. \ref{fig1} draws the  physically accessible and  physically inaccessible entanglement  $\rho_{AB_{\text{I}}}$ and   $\rho_{AB_{\text{II}}}$    as a function of the normalized distance ($r/r_{h}$) for the different constant $D_{HH}$ with $\alpha=\pi/4$.  As depicted in Fig. \ref{fig1}(a), it is interesting to find that: (i) the physically accessible   entanglement  $C(\rho_{AB_{\text{I}}})$ firstly decreases and then increases with the increasing normalized distance $r/r_{h}$, and it eventually converges on the fixed value of 1, which represents the maximal entanglement. (ii) \textcolor{black}{The extreme values of entanglement correspond to the peaks of Hawking radiation observed outside the event horizon, occurring within the range $r/r_{h}\in [1.43,1.5)$. This range represents the effective size of the black hole and "the peak" denotes the location where most Hawking emissions originate.} That is,
 the minimal entanglements for different $D_{HH}$   corresponds to the larger Hartle-Hawking temperature $T_{HH}$ resulting in the strongest Hawking radiation.
 Actually, the peaks of Hawking radiation can be obtained by solving $\partial_{r} T_{HH}|_{r=r_{\text {peak }}}=0$, to satisfy $D_{HH} \ge D_{c}\simeq 23.03$ for guaranteeing the real temperature, and $r_{\mathrm{peak}}$ represents the position at which the maximum temperature occurs. As a result, we have
\begin{equation}
    \begin{aligned}D_{HH}=&\frac{63-50r_{\mathrm{peak}}/r_{h}-(r_{\mathrm{peak}}/r_{h})^{2}}{8(r_{\mathrm{peak}}/r_{h}-3/2)}\\&+18\left(\frac{r_\mathrm{peak}}{r_h}-\frac32\right)\ln\frac{r_\mathrm{peak}}{r_h}.\end{aligned}
\end{equation}
The local temperature attains its maximum at a finite distance from the event horizon, with this distance exhibiting a monotonically increasing relationship as the parameter $D_{HH}$ is elevated. Notably, the local temperature peaks within the interval $1.43r_{h}\lesssim r_{\mathrm{peak}}<1.5r_{h}$ \cite{52}.
  (iii) The larger   $D_{HH}$, the more entanglement loss. Regardless of   $D_{HH}$, the maximum value always appears at $1.43\lesssim r/r_{h}<1.5$.

\begin{figure}[t]
    \centering
    \includegraphics[width=8cm]{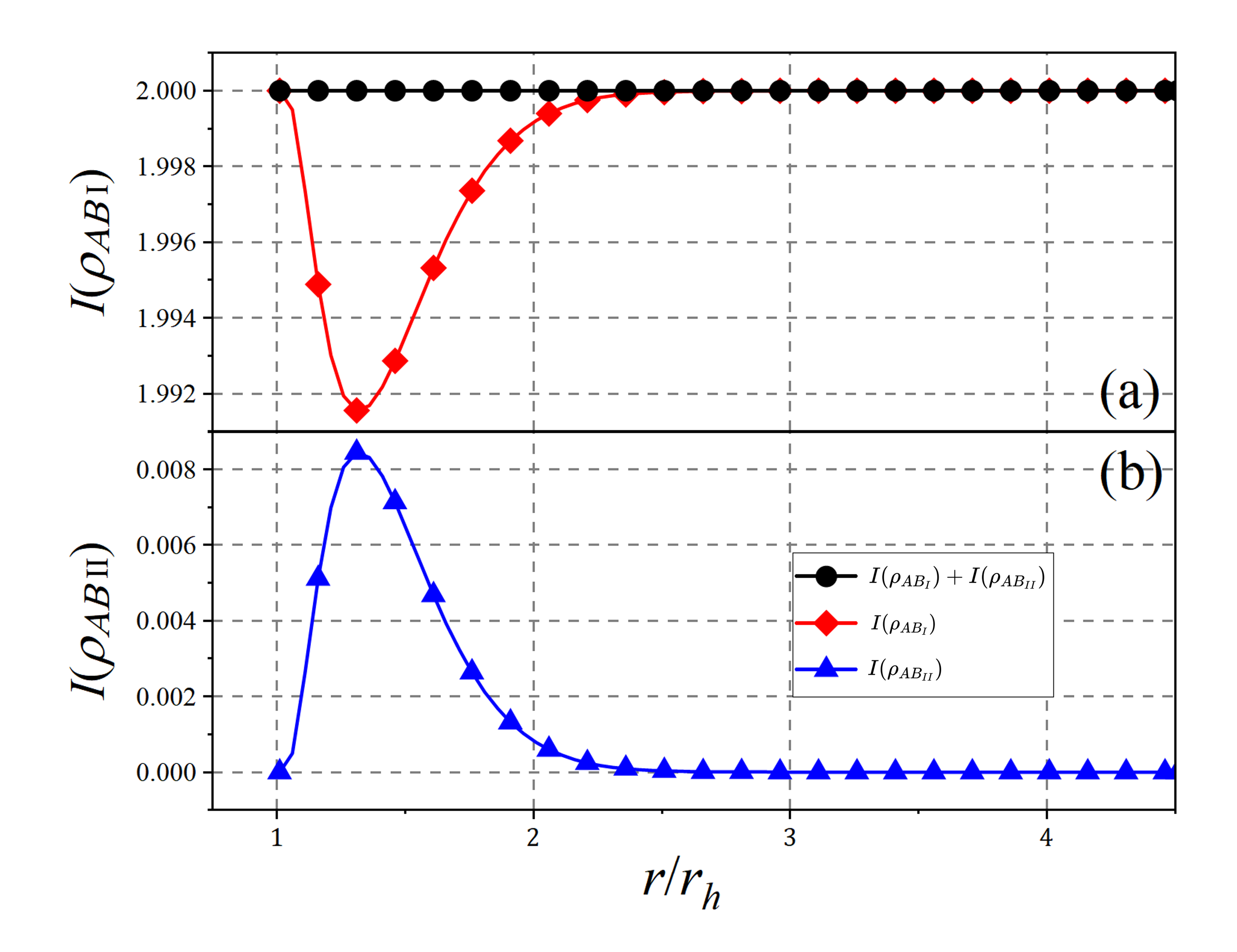}
    \caption{The mutual information as functions of the normalized distance $r/r_{h}$ for different states $\rho_{AB_{\text{I}}}$ and $\rho_{AB_{\text{II}}}$. Graph (a): the mutual information $I(\rho_{AB_{\text{I}}})$ of physically accessible region, and Graph (b): the mutual information $I(\rho_{AB_{\text{II}}})$ of physically inaccessible region. the state's parameter $\alpha= {\pi}/{4}$ and $D_{HH}=23.03$ are set in the plots.}
    \label{fig2}
\end{figure}

 By contrast,   the entanglement  $C(\rho_{AB_{\text{II}}})$ in the physically inaccessible regions, firstly increases and then decreases with the growth of  $r/r_{h}$, which are nearly inverse with  the entanglement in  the physically accessible region, as shown in Fig. \ref{fig1}(b).
 This can be interpreted by  that the Hawking effect associated with black holes induces redistribution of entanglement in the current architecture. Specifically, the strong Hawking effect diminishes bipartite entanglement in the physically accessible region while enhancing it in the physically inaccessible region.

To further capture quantum characteristics and interpret  information paradox in the black hole, we resort to examine the information redistribution by means of total correlation, i.e., mutual information.
Fig. \ref{fig2} offers the mutual information of the physically accessible region $I(\rho_{AB_{\text{I}}})$, the mutual information of the physically inaccessible region $I(\rho_{AB_{\text{II}}})$, and the total amount of $I(\rho_{AB_{\text{I}}})$ and $I(\rho_{AB_{\text{II}}})$. It is evident that, the dynamics of the mutual information is similar with those of entanglement, within the physically accessible region, mutual information initially increases and then decreases with the normalized distance, while in the physically inaccessible region, mutual information first decreases and subsequently increases. That implies the information from  the physically accessible region  flows toward the physically inaccessible regions and subsequently returns to the physically accessible regions, which suggests that the global information is redistributed between the physically accessible and inaccessible regions. Moreover, an intriguing phenomenon emerges: $I(\rho_{AB_{\text{I}}})$ and $I(\rho_{AB_{\text{II}}})$ are completely anti-correlated, and their total amount remains constant, verifying that the distribution of information adheres to the complimentary relationship. For example, $I(\rho_{AB_{\text{I}}})+I(\rho_{AB_{\text{II}}})=2$  is hold for \textcolor{black}{$\alpha=\pi/4$}.
Based on the above analysis, it is claimed that the obtained result supports the information-conservation principle, and  will benefit us better understand the information paradox in the framework of black holes as well.

\begin{figure}[t]
    \centering
    \includegraphics[width=8cm]{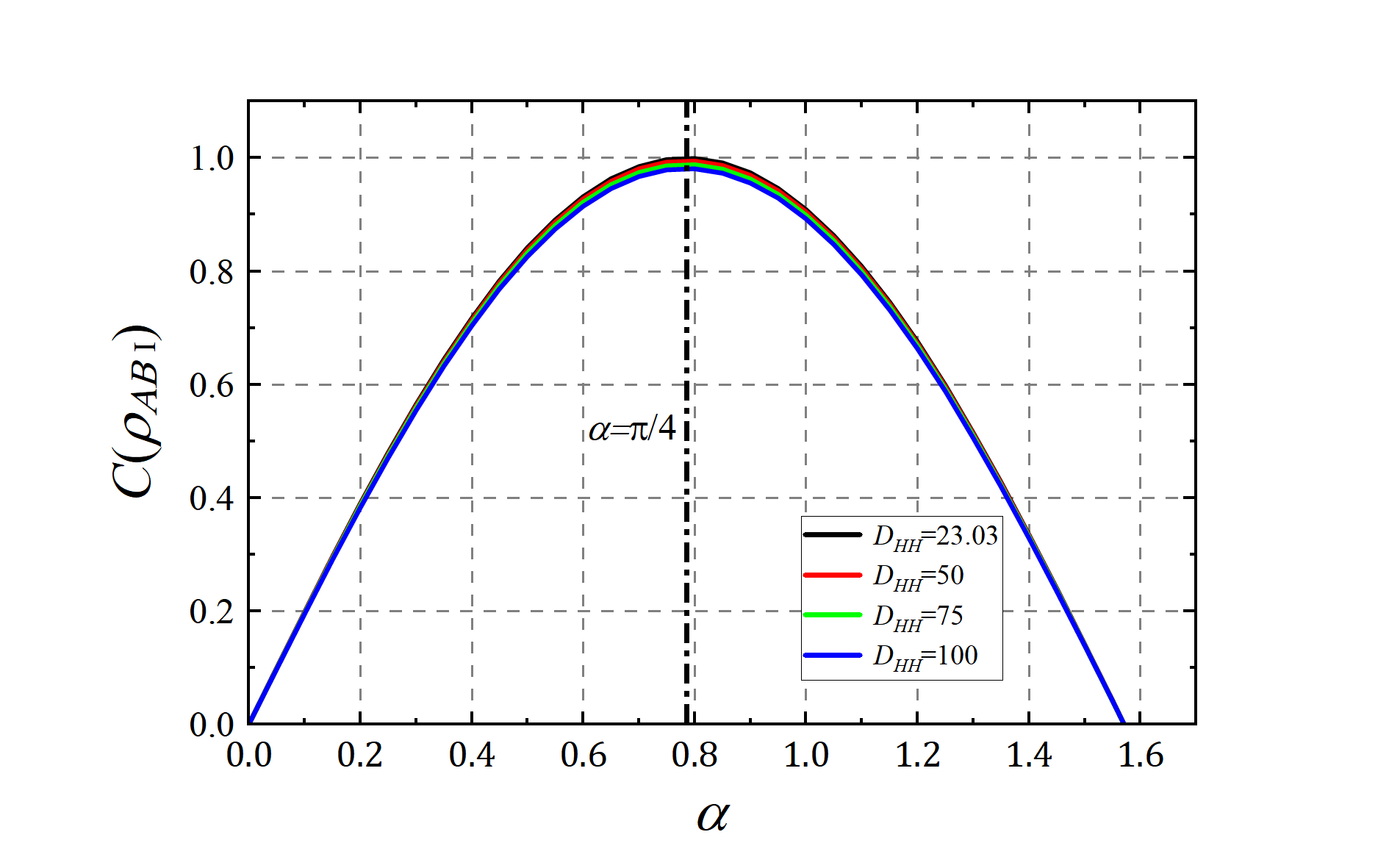}
    \caption{The bipartite entanglement $C(\rho_{AB_{\text{I}}})$ of physically accessible region as functions of state's parameter $\alpha$ for different $D_{HH}$. Here, $r/r_h=1.43$ is set. The black dashed line denotes that the entanglement reaches its maximum value at $\alpha=\pi/4$ .}
    \label{fig3}
\end{figure}

\begin{figure*}
    \centering
    \includegraphics[width=7cm]{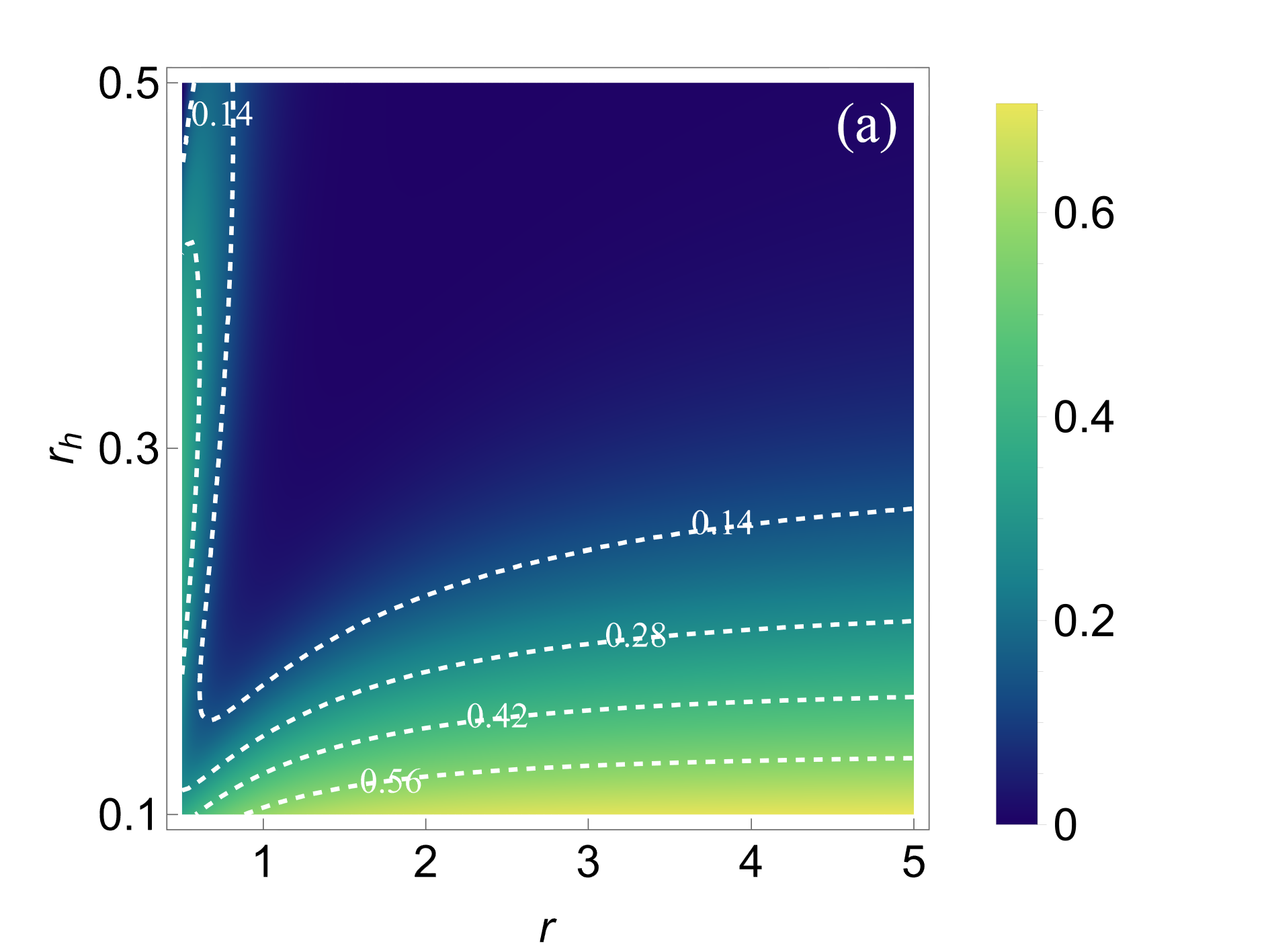}
    \includegraphics[width=7cm]{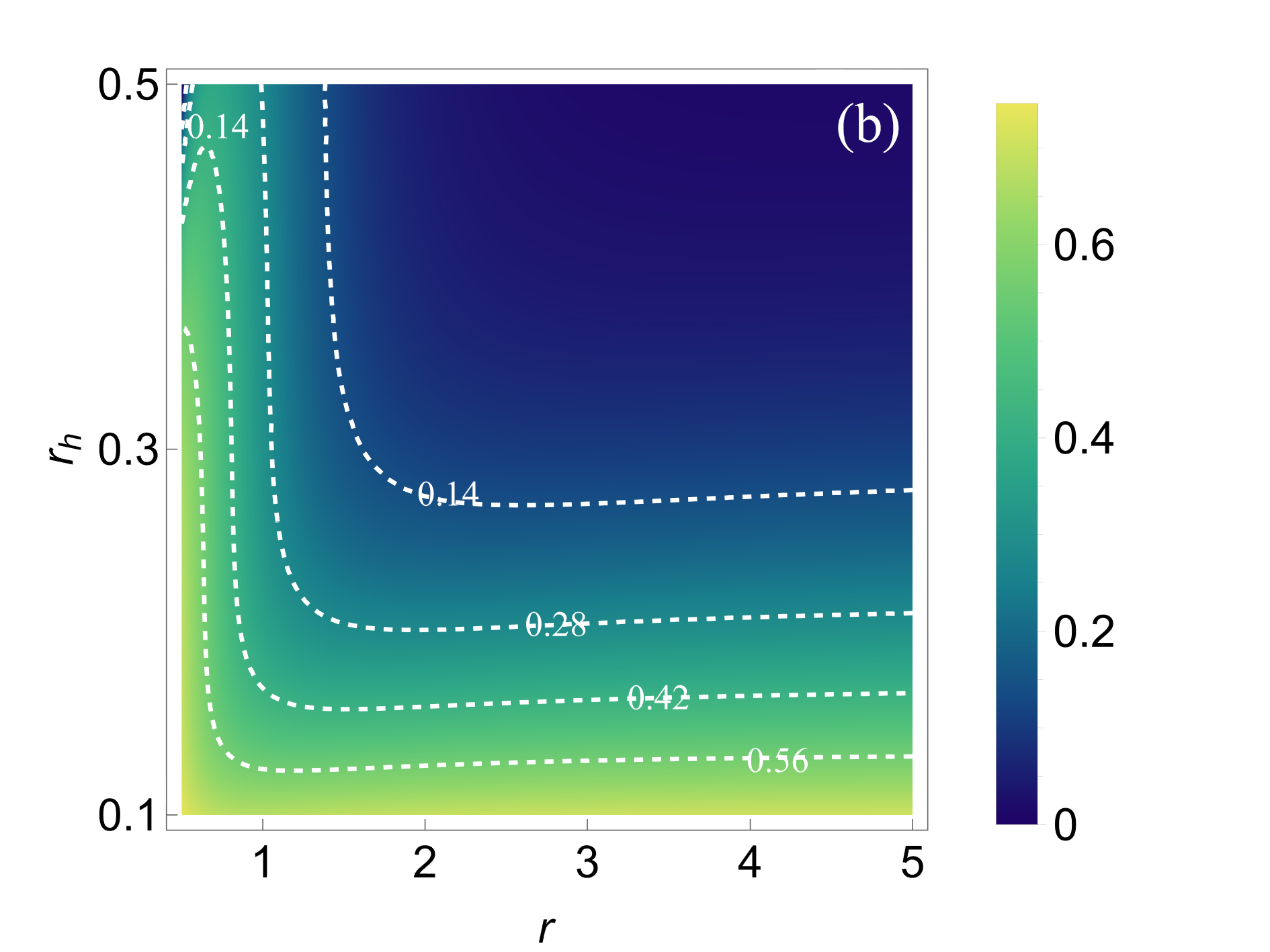}
    \includegraphics[width=7cm]{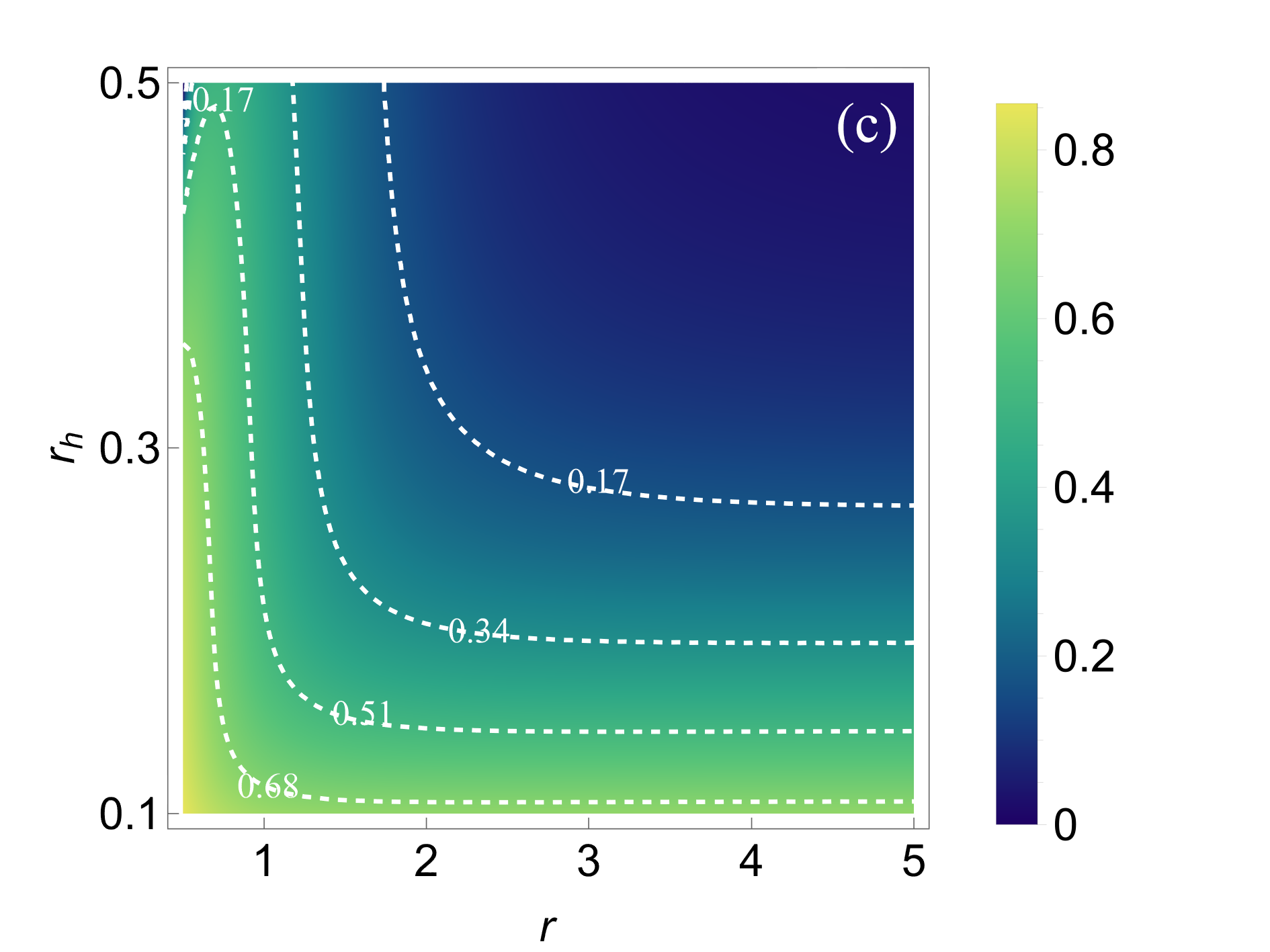}
    \includegraphics[width=7cm]{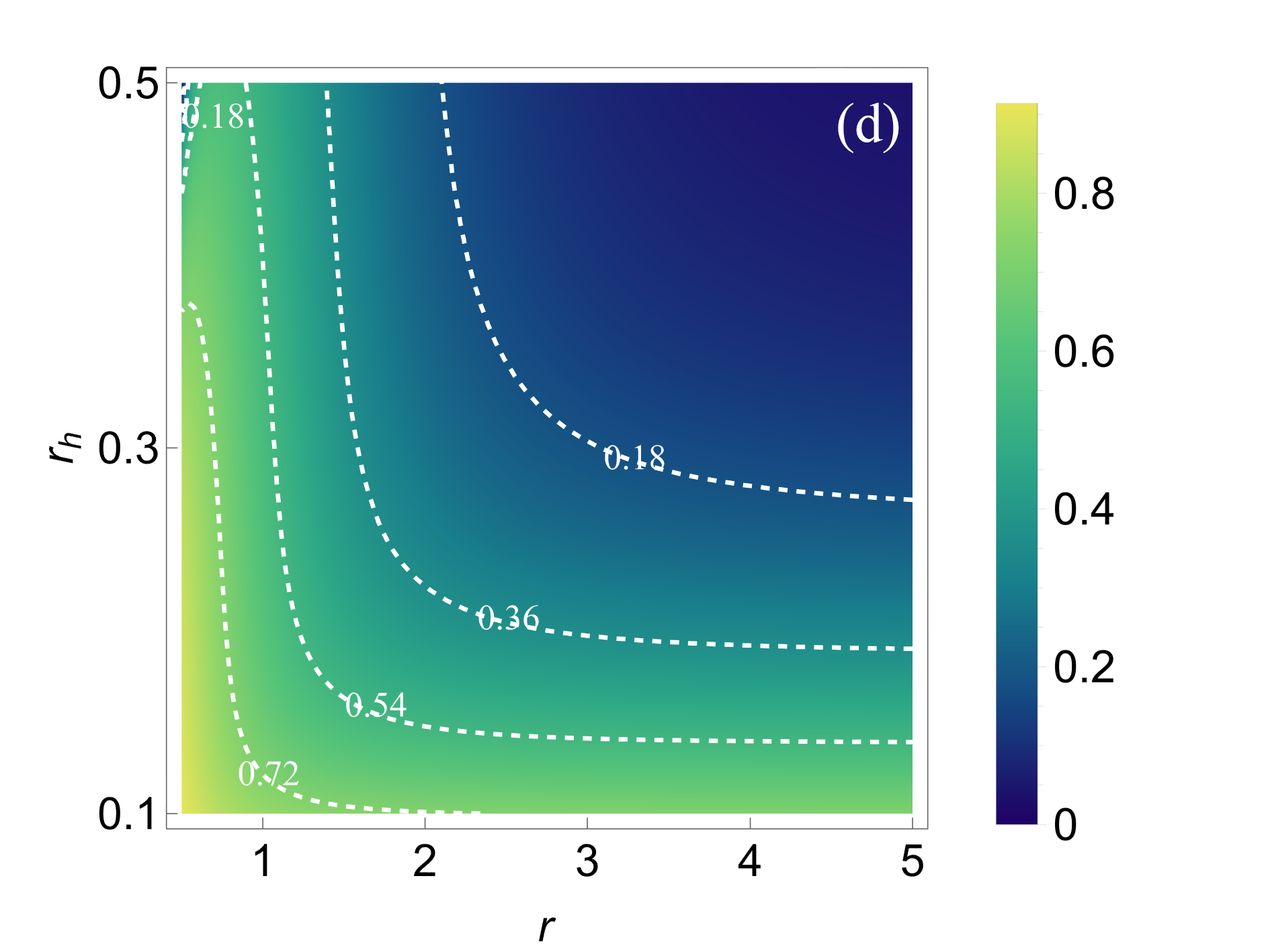}
    \caption{The uncertainty in the Hartle-Hawking vacuum as a function of the distance from the center of the black hole ($r$) and the event horizon size ($r_{h}$). Graph (a):  $D_{HH}=23.03$, Graph (b): $D_{HH}=50$, Graph (c): $D_{HH}=75$, and Graph (d): $D_{HH}=100$.   $\alpha=\arccos 0.7$ is chosen for all plots.}
    \label{fig4}
\end{figure*}

In addition,  we plot the concurrence of $\rho_{AB_{\text{I}}}$  versus  the state's parameter $\alpha$ for different $D_{HH}$ in Fig. \ref{fig3}. Following the figure, the entanglement initially increases and then decreases with the growth of $\alpha$. Obviously, the entanglement will reach its maximum value at the state's parameter of \textcolor{black}{$\alpha=\pi/4$}, and is also symmetric with \textcolor{black}{$\alpha=\pi/4$}. Besides, the entanglement decreases correspondingly as the $D_{HH}$ increases, which essentially is in agreement with the conclusion made before.

Next, we turn to take into account the measurement uncertainty, which is nontrivial for perspective quantum information processing in the current scenario. To explore the effect of different horizon radiuses on the uncertainty, we plot  the uncertainty in the physically accessible region as functions of the parameters $r_{h}$ and $r$ for various $D_{HH}$, as displayed in Fig. \ref{fig4}.
From the figure, we obtain that: (i) in general,  the uncertainty firstly increases and then decreases with the increasing distance $r$. While the uncertainty will monotonically decrease as $r_h$ increases.
And the magnitude of the uncertainty will saturate into zero for $r,r_h\rightarrow\infty$, in which case the system will not suffer from any Hawking radiation. (ii) The region of the non-zero uncertainty
becomes larger as $D_{HH}$ increases, implying $D_{HH}$  will inflate the uncertainty in a large degree. (iii)  By comparing Figs. \ref{fig1} and \ref{fig4}, the uncertainty and entanglement are inversely correlated for the increasing $D_{HH}$. That is, the stronger entanglement will induce the smaller measurement uncertainty, and vice versa.

\section{Discussion and conclusions}
In this study, we investigate the Hawking effect of the Dirac field in Schwarzschild black holes on quantumness of the quantum atmosphere.  Specifically, we analyze entanglement dynamics with the normalized distance, and it is revealed that the entanglement is redistributed in physically accessible and  inaccessible  regions. Explicitly, the entanglement in the physically accessible region first increases and then decreases with the increasing normalized distance, while the entanglement shows the opposite variation in the physically inaccessible region. Remarkably, the entanglement exhibits significant extremum characteristics within $r/r_{h}\in [1.43,1.5)$, corresponding to \textcolor{black}{the effective size} of the black hole and closely relating to the peaks of Hawking radiation in the quantum atmosphere.
\textcolor{black}{Furthermore, when considering higher-dimensional Schwarzschild black holes, the event horizon becomes smaller for a given mass \cite{53}, resulting in a more ``compact'' geometric structure. As the number $D+1$ of spacetime dimensions increases, the effective size of the black hole gradually approaches the event horizon \cite{54}. In this case, the extremal features of entanglement still exist, with the location of the extremum moving closer to the horizon as the dimensionality increases. For charged black holes (e.g., Reissner-Nordstr\"{o}m black holes), the size of the quantum atmosphere is defined by the wavelength of the Hawking particle, as the charge-to-mass ratio ($Q/M$) increases, the location of the quantum atmosphere moves outward \cite{55}. Correspondingly, the extremal features of entanglement also shift away from the event horizon. In the extremal limit ($Q \rightarrow M$), the location of the quantum atmosphere eventually diverges to infinity, and the temperature also decreases to zero in the same limit. The extremal feature of entanglement may disappear. }

By virtue of the mutual information, we demonstrate that  the information from
the physically accessible region  flows toward the physically inaccessible regions and then goes back the physically accessible ones. We also identify the complementary relationship of the information distribution.
In addition, the uncertainty of the model is examined as well, and it turns out that  the uncertainty is inversely correlated with the entanglement. Moreover,
$D_{HH}$ has a significant destructive effect on the entanglement of the system.  With these in mind, it is argued that our investigations offer insights into the quantumness and uncertainty in the quantum atmosphere, and promote us better understand the information paradox in the background of black holes.

\section*{Acknowledgements} 
This work was supported by the National Science Foundation of China (Grant nos. 12475009, 12075001, and 62471001), Anhui Provincial Key Research and Development Plan (Grant No. 2022b13020004), Anhui Province Science and Technology Innovation Project (Grant No. 202423r06050004) and Anhui Provincial University Scientific Research Major Project (Grant No. 2024AH040008).
	
\newcommand{\PRL}{\emph{Phys. Rev. Lett.} }
\newcommand{\RMP}{\emph{Rev. Mod. Phys.} }
\newcommand{\PRA}{\emph{Phys. Rev. A} }
\newcommand{\PRB}{\emph{Phys. Rev. B} }
\newcommand{\PRE}{\emph{Phys. Rev. E} }
\newcommand{\PRD}{\emph{Phys. Rev. D} }
\newcommand{\APL}{\emph{Appl. Phys. Lett.} }
\newcommand{\NJP}{\emph{New J. Phys.} }
\newcommand{\JPA}{\emph{J. Phys. A} }
\newcommand{\JPB}{\emph{J. Phys. B} }
\newcommand{\OC}{\emph{Opt. Commun.} }
\newcommand{\PLA}{\emph{Phys. Lett. A} }
\newcommand{\EPJD}{\emph{Eur. Phys. J. D} }
\newcommand{\NP}{\emph{Nat. Phys.} }
\newcommand{\NC}{\emph{Nat. Commun.} }
\newcommand{\EPL}{\emph{Europhys. Lett.} }
\newcommand{\AoP}{\emph{Ann. Phys.} }
\newcommand{\ADP}{\emph{Ann. Phys. (Berlin)} }
\newcommand{\QIC}{\emph{Quantum Inf. Comput.} }
\newcommand{\QIP}{\emph{Quantum Inf. Process.} }
\newcommand{\CPB}{\emph{Chin. Phys. B} }
\newcommand{\IJTP}{\emph{Int. J. Theor. Phys.} }
\newcommand{\IJMPB}{\emph{Int. J. Mod. Phys. B} }
\newcommand{\PR}{\emph{Phys. Rep.} }
\newcommand{\SR}{\emph{Sci. Rep.} }
\newcommand{\LPL}{\emph{Laser Phys. Lett.} }
\newcommand{\OEE}{\emph{Opt. Exp.} }
\newcommand{\IJQI}{\emph{Int. J. Quantum Inf.} }
\newcommand{\EPJC}{\emph{Eur. Phys. J. C} }

\newpage
\bibliographystyle{plain}

\end{document}